\documentclass[epjCONF,columns]{svjour} 
\usepackage{graphics}
\usepackage[varg]{txfonts} % Times fonts
\usepackage[latin1]{inputenc}

\usepackage{ads_refs}

\usepackage[english]{babel}
\usepackage{graphicx}
\usepackage{verbatim}
\usepackage{multirow}

\begin{document}
\title{The rate of stellar tidal disruption flares from SDSS data}
\author{ Sjoert van Velzen\inst{1}\fnmsep\thanks{\email{s.vanvelzen@astro.ru.nl}} \and Glennys R. Farrar\inst{2,3}}
\institute{Department of Astrophysics/IMAPP, Radboud University, P.O. Box 9010, 6500 GL Nijmegen, The Netherlands \and 
Center for Cosmology and Particle Physics, New York University, NY 10003, USA \and
Department of Physics, New York University, NY 10003, USA}
\abstract{
We have searched for flares due to the tidal disruption of stars by supermassive black holes in archival Sloan Digital Sky Survey (SDSS) multi-epoch imaging data. Our pipeline takes advantage of the excellent astrometry of SDSS to separate nuclear flares from supernovae. The 10 year baseline and the high cadence of the observations facilitate a clear-cut identification of variable active galactic nuclei. We found 186 nuclear flares, of which two are strong stellar tidal disruption flare (TDF) candidates. To compute the rate of these events, we simulated our entire pipeline to obtain the efficiency of detection for a given light curve. We compute a model-independent upper limit to the TDF rate of $\dot{N}<3\times 10^{-4}\,{\rm yr}^{-1}{\rm galaxy}^{-1}$ (90\% CL). Using a simple model to extrapolate the observed light curve forward and backward in time, we find our best-estimate of the rate: $\dot{N}=3_{-3}^{+5}\times 10^{-5}\,{\rm yr}^{-1}{\rm galaxy}^{-1}$.
} 

\maketitle

\section{Introduction}
A star that comes too close to a massive black hole is torn apart by tidal gravity forces, yielding a stellar tidal disruption flare (TDF). How many stars per galaxy per year suffer this fate is currently not well constrained (see \cite{Alexander12} for a recent review). Based on the nuclear density profiles of nearby elliptical galaxies, Wang and Merritt \cite{Wang_Merritt04} compute a rate of $\sim 10^{-4}~{\rm yr}^{-1}$, with 1~dex scatter between different galaxies of similar mass. 

Recently, two TDFs were found in the Sloan Digital Sky Survey (SDSS) multi-epoch imaging data \cite[hereafter paper~I]{vanVelzen10}. The systematic nature of this search allows for a relatively straightforward computation of the rate of these events. In this proceeding, we shall first summarize the search of paper~I, followed by a derivation of the TDF rate. Full details on rate analysis will be given in a forthcoming publication.

\subsection{Summary of SDSS nuclear flare search}\label{sec:sum}
The search for TDFs (paper~I) was conducted in Stripe~82. 
Nuclear flares in galaxies are found using two steps: \emph{(i)} a series of catalog cuts to select flaring galaxies, followed by \emph{(ii)} careful difference imaging to measure the angular distance between the flare and the host. The catalog cuts selected galaxies with a flux increase of 10\% or more, detected at the 7$\sigma$-level. A sample of 186 nuclear flares was selected based on the distance between the center of the host and the center of the flare in the difference image ($d<0.2"$). 
After removing galaxies that fall inside the photometric QSO locus and removing galaxies with additional variability, two flares remained: TDE1 and TDE2. Additional analysis and follow-up observations show that these flares are best explained as stellar tidal disruption events (paper~I). 

\section{Analysis}
The number of detected flares in a variability survey that targets galaxies is given by
\begin{equation}\label{eq:ratefull}
N_{\rm TDF} = \tau \, \sum_i \epsilon_i \dot{N}_i  \quad
\end{equation}
where $\dot{N}_i$ and $\epsilon_i$ are the flare rate and detection efficiency for the $i$th monitored galaxy, and $\tau$ is the survey time. The rate of TDFs is expected to depend only weakly on black hole mass \cite{Magorrian_Tremaine99}. Furthermore, as we shall show below, our search is sensitive to a relatively narrow range of black hole masses. We can therefore simplify Eq. \ref{eq:ratefull} using $\dot{N}_i=\dot{N}$, a galaxy-independent rate, to find
\begin{equation}\label{eq:rate}
\dot{N} = \frac{N_{\rm TDF}} {N_{\rm gal} \tau \, \epsilon} \quad.
\end{equation} 
Here we defined we mean overall efficiency $\epsilon \equiv N^{-1}\sum_i^N\epsilon_i$. For the TDF search in Stripe~82 we set $\tau=7.6~{\rm yr}$, starting in the year 2000, $N_{\rm TDF}=2$, and $N_{\rm gal}=1.6 \times 10^6$, i.e., all galaxies with a photometric redshift that are outside the QSO locus. Finding $\dot{N}$ thus boils down to computing the mean overall efficiency, $\epsilon$. 

As discussed in sec. \ref{sec:sum}, the detection pipeline of paper~I consists of two stages: the catalog cuts and the difference imaging. Since the catalog cuts are applied to the Petrosian flux \cite{Stoughton02} of the galaxy, computing the probability that a simulate light curve passes these cuts is trivial: one simply adds the flare flux to the galaxy flux and reruns the catalog cuts. To estimate the detection probability of the difference imaging pipeline we selected 1400 random galaxies in uniform magnitude bins and inserted point sources at the center of their images. The detection probability as a function of flare and host magnitude follows from the number of detected point sources in each magnitude bin.

\begin{figure}
\centering
\includegraphics[trim=5mm 50mm 15mm 80mm, clip, width=.46 \textwidth]{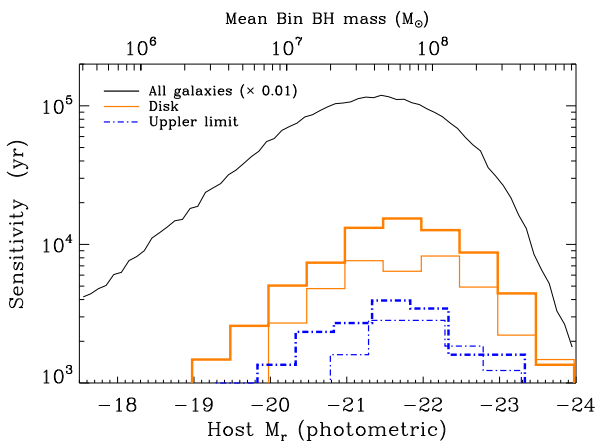}
\caption{Sensitivity of our search (i.e., the denominator of Eq. \ref{eq:rate}, $N_{\rm gal} \tau  \epsilon$), in bins of absolute magnitude of the host (thin and thick lines for TDE1 and TDE2, respectively). The mean black hole mass in each bin is indicated on the upper axis. We also show the parent galaxy sample by setting $\epsilon=1$ (thin black line). Our analysis is most sensitive to galaxies with $M_r = -21.5$ or $M_{\rm BH} \sim 5 \times 10^7 M_\odot$. 
}\label{fig:rate}
\end{figure}

With the detection probabilities measured, we now proceed to compute the overall efficiency ($\epsilon$ in Eq. \ref{eq:rate}). For a given galaxy we first draw a location for the start of the flare for a uniform distribution. We then add the flare flux to the Petrosian flux and check if this galaxy would pass our catalog cuts. In the final step, we simulate the detection of this flare (for at least three nights in the $u$, $g$, and $r$ bands) using the probability of detection for the flare and host magnitude. After repeating this process for a sufficiently large sample of galaxies, the overall efficiency follows from the number of flares detected by the model pipeline over the number of simulated flares.
% (convergence of this method is confirmed in sec. \ref{sec:rate}). 
Because the flares are inserted into the observed galaxy light curves, this method fully takes into account the inhomogeneous cadence and varying data quality of Stripe~82. 

%\subsection{Simulated light curves}
The detection efficiency will obviously depend on the flare's luminosity and duration. 
%-- e.g., a long, bright flare will be above detection threshold long after the peak and thus is more readily detected with a given set of observations. 
Because the SDSS data of the two TDFs of paper~I does not completely cover the time they are detectable (i.e., above the flux limit), we have to extrapolate the observed light curve forward and backward in time. 
Since our optical observations probe the Rayleigh-Jeans regime of the SED, we adopt $L_{\rm TDF}\propto t^{-5/12}$ \cite{Lodato11}. We fit the observed light curve to $F\propto (t-t_D)^{-5/12}$ to find the time of disruption, $t_D$. To fix the normalization of the light curve, we shall assume that the luminosity of the flare is proportional to the Eddington luminosity, $L_{\rm TDF}\propto M_{\rm BH}$. An estimate of the black hole mass ($M_{\rm BH}$) is obtained from the galaxy luminosity using the black hole-bulge mass relation \cite{HaringRix04} and the dynamical mass-to-light ratio from the fundamental plane, yielding $M_{\rm BH} \propto  L_{\rm bulge}^{1.3}$, with $L_{\rm bulge}$ the bulge luminosity. 
In summary, the expression for the model light curve of a flare in the $i$-th galaxy is
\begin{equation}\label{eq:modellc}
M_{i}(t) = M_{\rm TDE1,2}(t) + 1.3 \, (M_{i,{\rm bulge}}-M_{{\rm TDE1,2},{\rm bulge}})\quad 
\end{equation}
with $M_{{\rm TDE1,2}}(t)$ the (extrapolated) light curves of TDE1,2. We use the galaxy photometric redshift \cite{oyaizu08} to convert between apparent and absolute magnitudes.  

The light curves we simulate are 300 days long, extrapolated back in time to five days after the disruption, $t_0 = (5 + t_D)$ (we do not extrapolate back further because at very early time the power-law scaling is not appropriate). 
The final efficiency is quite insensitive to the light curve length or $t_0$ because we required at least three detections of the flare within the same season.

Finally, we also compute a model-independent upper limit to the TDF rate by using only the observed fraction of the light curve of TDE1,2.

\section{Results}\label{sec:rate}
Using the model light curves presented in the previous section (Eq. \ref{eq:modellc}), we injected flares scaled to either TDE1 or TDE2 to find the overall efficiency for detecting these flares. We use the average of these two overall efficiencies to obtain the TDF rate (Eq. \ref{eq:rate}); the results are shown in Table \ref{tab:model}. The uncertainties on the rate are given by the 90\% confidence level (CL) for Poisson statistics; for the upper limit on the rate we use $N_{\rm TDF}<5.3$, the 90\% CL upper limit if two events are detected.

In Fig. \ref{fig:rate} we show the sensitivity of our pipeline as a function of host luminosity. The sensitivity to TDFs from faint $M_r<-20$ galaxies decreases faster than their observed number density, which is due to the scaling of the flare luminosity with black hole mass. Our search is most sensitive to galaxies in the range $-21<M_r<-22.5$. For each host we can estimate the central black hole mass from the bulge luminosity using the black hole mass-bulge mass relation, calibrated for the SDSS $r$-band \cite{Tundo07}. We find that our analysis is most sensitive to black holes in the mass range $2\times 10^7~M_\odot$ to $3 \times 10^8~M_\odot$.

The TDF rate we derive with the simplified model light curve is below some of the theoretical predictions that have been published \cite{Wang_Merritt04}, but consistent with the rate of soft X-ray flares from quiescent galaxies \cite{donley02}.  Our strictly observational upper bound is not strong enough to rule out any current predictions.

\begin{table}
\centering
%\large
\caption{Measured rate of optical stellar tidal disruption flares} \label{tab:model}
\begin{tabular}{l c c}
\hline\hline
{} & {Light curve}  & Rate \\ 
 {} & {($F_{\rm TDF}\propto t^{p}$)} & {(${\rm galaxy}^{-1}~{\rm yr}^{-1}$)} \\ 
\hline\hline
  Upper limit & Observed & $<3\times 10^{-4}$ \\  
  Disk model & $p=-\frac{5}{12}$ &  $3_{-3}^{+6}\times 10^{-5}$ \\ 
%\hline
\end{tabular}
\end{table}


\begin{thebibliography}{9}

\bibitem{Alexander12}
T.~{Alexander}, arXiv e-prints (2012) \texttt{1210.0582} 

\bibitem{Wang_Merritt04}
J.~{Wang}, D.~{Merritt}, \apj ~\textbf{600}, 149 (2004)  

\bibitem{vanVelzen10}
S.~{van Velzen}, G.R. {Farrar}, S.~{Gezari}, N.~{Morrell}, D.~{Zaritsky},
  L.~{{\"O}stman}, M.~{Smith}, J.~{Gelfand}, A.J. {Drake}, \apj ~\textbf{741},
  73 (2011)

\bibitem{Magorrian_Tremaine99}
J.~{Magorrian}, S.~{Tremaine}, \mnras ~\textbf{309}, 447 (1999)

\bibitem{Lodato11}
G.~{Lodato}, E.M. {Rossi}, \mnras ~\textbf{410}, 359 (2011)

\bibitem{Stoughton02}
C.~{Stoughton} {et~al.}, \aj ~\textbf{123}, 485 (2002)


\bibitem{HaringRix04}
N.~{H{\"a}ring}, H.~{Rix}, \apjl ~\textbf{604}, L89 (2004)

%\bibitem{bender92}
%R.~{Bender}, D.~{Burstein}, S.M. {Faber}, \apj ~\textbf{399}, 462 (1992)

\bibitem{oyaizu08}
H.~{Oyaizu}, M.~{Lima}, C.E. {Cunha}, H.~{Lin}, J.~{Frieman}, E.S. {Sheldon},
  \apj ~\textbf{674}, 768 (2008)

\bibitem{Tundo07}
E.~{Tundo}, M.~{Bernardi}, J.B. {Hyde}, R.K. {Sheth}, A.~{Pizzella}, \apj
  ~\textbf{663}, 53 (2007)

\bibitem{donley02}
J.L. {Donley}, W.N. {Brandt}, M.~{Eracleous}, T.~{Boller}, \aj ~\textbf{124},
  1308 (2002)


\end{thebibliography}
\end{document}